\documentclass[12pt]{iopart}

\usepackage{graphicx}
\usepackage{bm}
\usepackage{epsfig}
\begin{document}
\bibliographystyle{unsrt}

\title[Computation of percolation critical polynomials]{The computation of generalized percolation critical polynomials by the deletion-contraction algorithm}


\author{Christian R. Scullard}

\address{Lawrence Livermore National Laboratory, Livermore CA 94550, USA}

\begin{abstract}
Although every exactly known bond percolation critical threshold is the root in $[0,1]$ of a lattice-dependent polynomial, it has recently been shown that the notion of a critical polynomial can be extended to any periodic lattice. The polynomial is computed on a finite subgraph, called the base, of an infinite lattice. For any problem with exactly known solution, the prediction of the bond threshold is always correct, regardless of the base chosen. For unsolved problems, the polynomial is referred to as the generalized critical polynomial and provides an approximation that becomes more accurate with increasing number of bonds in the base, appearing to approach the exact answer. The polynomials are computed using the deletion-contraction algorithm, which quickly becomes intractable by hand for more than about 18 bonds. Here, I present generalized critical polynomials calculated with a computer program for bases of up to 36 bonds for all the Archimedean lattices, except the kagome which was considered in an earlier work. The polynomial estimates are generally within $10^{-5}$ to $10^{-7}$ of the numerical values, but the prediction for the $(4,8^2)$ lattice, though not exact, is not ruled out by simulations.
\end{abstract}

\maketitle
\section{Introduction}
\label{sec:intro}
\begin{figure}
\begin{center}
\includegraphics{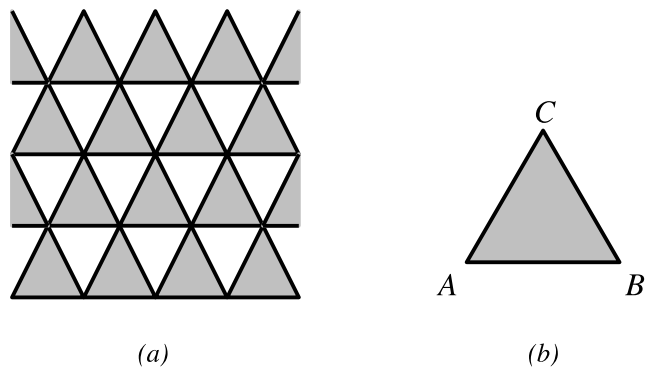}
\caption{a) A self-dual 3-uniform hypergraph. The shaded triangles can be any configuration of sites and bonds lying within the three boundary vertices; b) the critical point is given by (\ref{eq:allnone}).}
\label{fig:selfdual}
\end{center}
\end{figure}

Percolation \cite{Stauffer} is one of the simplest random processes taking place on a lattice. Nonetheless, since its introduction over fifty-five years ago \cite{BroadbentHammersley}, it has continued to provide physicists and mathematicians with an array of fascinating problems (see \cite{Saleur1987,Pinson,Damron2011,Simmons2007,Kleban2003,Beffara} for a small sampling) and has inspired many new mathematical techniques \cite{Cardy92,Smirnov,Schramm2001}. Given an infinite lattice, $L$, we declare each edge to be open with probability $p$ and closed with probability $1-p$. When $p$ is small, $L$ will be sparsely populated by small clusters of open bonds. When $p$ is near 1, we will have an infinite open mass with small pockets of closed bonds. In between this regime lies the critical threshold, $p_c$, which marks the transition from the unconnected phase to the phase containing an infinite open cluster. One of the most challenging problems in the field is the analytic determination of critical thresholds. Outside of one dimension and a narrow class of two-dimensional lattices \cite{Scullard06,Ziff06,Bollobas}, exact results remain elusive. In some cases, mathematically rigorous confidence intervals \cite{Riordan} and bounds \cite{MayWierman05,Wierman2003} have been proved and which are continually improving, but, for most lattices, critical probabilities are known only numerically \cite{ZiffSuding97,SudingZiff99,Parviainen,Ding2010,Feng08}. All solved two-dimensional lattices are formed from self-dual 3-uniform hypergraphs, such as the one shown in Figure \ref{fig:selfdual}a, where the shaded triangle in Figure \ref{fig:selfdual}b permits any configuration of sites and bonds as long as they lie within the three boundary vertices. For such graphs, the critical condition is given by \cite{Ziff06,ChayesLei,Bollobas}
\begin{equation}
 P(A,B,C)=P(\bar{A},\bar{B},\bar{C}) \label{eq:allnone}
\end{equation}
where $P(A,B,C)$ is the probability that all three corners are connected and $P(\bar{A},\bar{B},\bar{C})$ is the probability that none are connected. For bond percolation, for example, this condition allows one to find the inhomogeneous critical surface by assigning each bond in the shaded triangle a different probability. The homogeneous bond threshold is given by a polynomial in the probability $p$, of degree equal to the number of bonds in the triangle. In previous work \cite{Scullard08,Scullard10,Scullard11} it was shown that one can define a critical polynomial on any lattice that agrees with (\ref{eq:allnone}) for self-dual 3-uniform lattices, and provides an accurate approximation for unsolved problems. This polynomial is a kind of graph invariant \cite{Scullard11-2} with similarity to the Tutte polynomial \cite{Bollobasbook} in that it may be computed by the deletion-contraction algorithm. In fact, deletion-contraction is how the generalized critical polynomial is defined, as we will see in the following section. An infinite lattice may be partitioned by a finite subgraph which is tiled in a regular way to give the full lattice. The critical polynomial is defined on this subgraph, called the base, and is thus a property of a finite graph and its embedding into the infinite lattice. Using the example of the kagome lattice, it was demonstrated in \cite{Scullard11-2} that polynomials computed on bases of increasing numbers of bonds provide better approximations to the critical thresholds, with bases of $36$ bonds making predictions within $10^{-7}$ of the numerically determined critical threshold. 

\begin{figure}
\begin{center}
\includegraphics{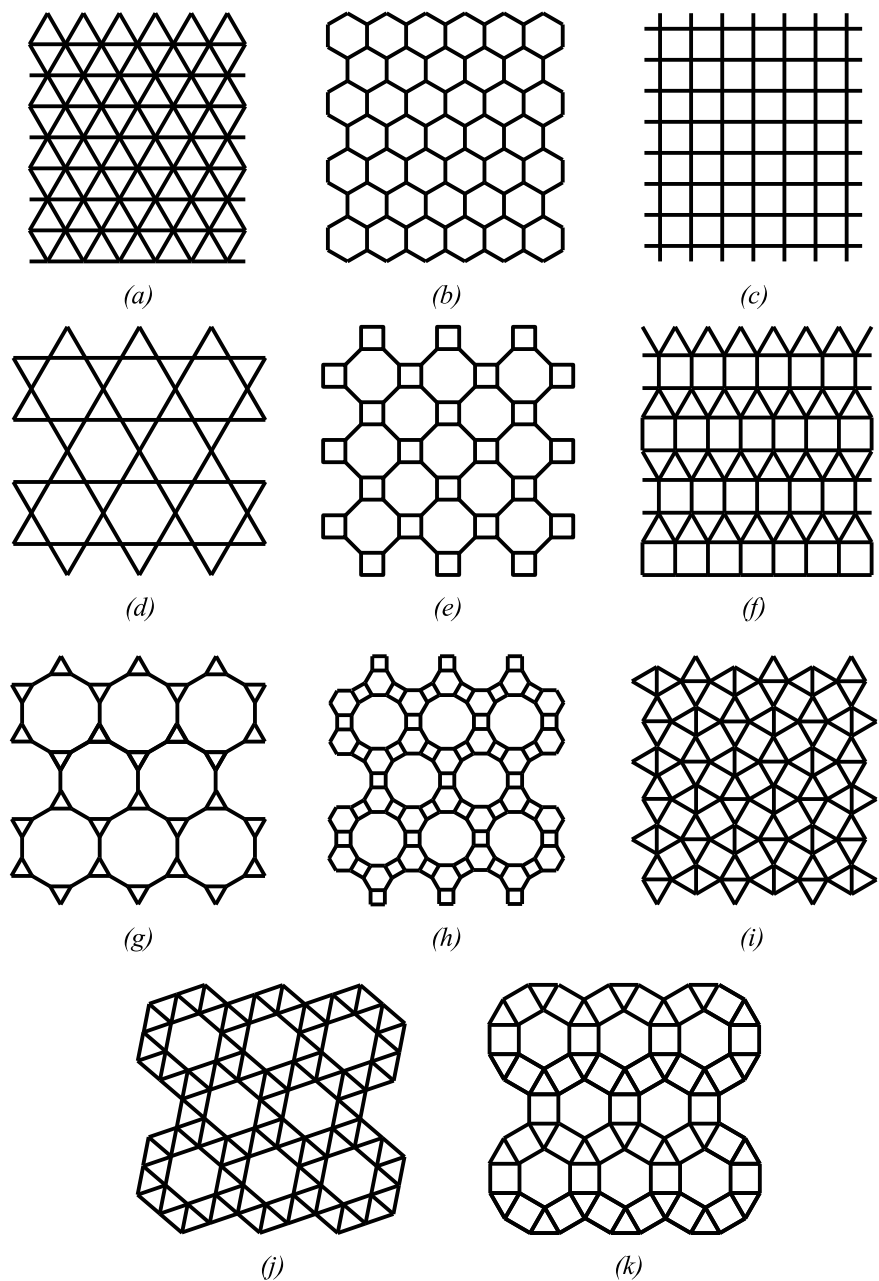}
\caption{The Archimedean lattices; a) triangular; b) hexagonal; c) square; d) kagome; e) $(4,8^2)$; f) $(3^3,4^2)$; g) $(3,12^2)$; h) $(4,6,12)$; i) $(3^2,4,3,4)$; j) $(3^4,6)$; k) $(3,4,6,4)$ .}
\label{fig:archimedean}
\end{center}
\end{figure}

In this paper, I use the computer program used in \cite{Scullard11-2} for the kagome lattice to find polynomials for the remaining unsolved Archimedean lattices (Fig. \ref{fig:archimedean}), providing further evidence for the conjecture that the estimates made by these critical thresholds approach the exact answer as the size of the base increases. I begin by reviewing the definition of the polynomial in the next section. In section \ref{sec:impl}, I describe in detail the operation of the program to automate its computation. The subsequent sections are devoted to reporting polynomials for the Archimedean lattices partitioned into different bases.
\section{Deletion-contraction algorithm}
\begin{figure}
\begin{center}
\includegraphics{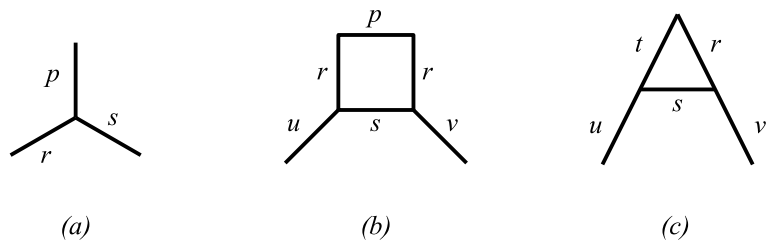}
\caption{Assignments of probabilities on a) the hexagonal lattice, b) the $(4,8^2)$ lattice, c) the martini-A lattice.}
\label{fig:hex_FE_A}
\end{center}
\end{figure}

Consider the hexagonal lattice of Figure \ref{fig:archimedean}b, with assignment of probabilities on the unit cell shown in Figure \ref{fig:hex_FE_A}a. The condition (\ref{eq:allnone}) then yields the critical surface
\begin{equation}
 H(p,r,s) \equiv p r s - pr -ps -rs +1=0 \ .
\end{equation}
As with all surfaces determined by (\ref{eq:allnone}), this expression is first-order in all its arguments. Consider the $(4,8^2)$ lattice of Figure \ref{fig:archimedean}e, with the assignment of probabilities in Figure \ref{fig:hex_FE_A}b. This lattice has an unsolved bond threshold as it does not fall into the class that may be found with (\ref{eq:allnone}). However, we generalize the notion of the critical polynomial on this lattice by assuming that its critical surface has the first-order property. Next, we note that contracting its $p$--bond by setting $p=1$ yields the martini-A lattice \cite{Scullard06}. This lattice is formed by substituting the generator in Figure \ref{fig:hex_FE_A}c for the shaded triangles in Figure \ref{fig:selfdual}a, and thus its critical surface is known exactly. It has previously been reported in many places \cite{Scullard10,ZiffScullard06,Wu06} so here we just denote it $A(r,s,t,u,v)$. Deleting the $p$--bond by setting $p=0$ gives the hexagonal lattice with some bonds doubled in series. Reduction to these two cases and the imposition of the first-order property leaves us with only one choice for the critical surface of the $(4,8^2)$ lattice, given by the deletion-contraction formula,
\begin{equation}
 \mathrm{FE}(p,r,s,t,u,v)=p A(r,s,t,u,v)+(1-p)H(s,ur,tv) \ . \label{eq:DC}
\end{equation}
Expanding this, and setting all probabilities equal gives the homogeneous polynomial,
\begin{equation}
 1- 4 p^3 - 2 p^4 + 6 p^5 - 2 p^6 = 0 \ , \label{eq:FE1}
\end{equation}
with solution on $[0,1]$, $p_c=0.67683519...$\ . The simulation result reported by Parviainen \cite{Parviainen} is $p_c^\mathrm{num}=0.67680232(63)$ (the brackets indicate the standard error on the last digits), which, although the difference is only $3.3 \times 10^{-5}$, easily rules out our estimate. However, we may extend the inhomogeneous probabilities over two unit cells. That is, we may consider a larger base. This is shown in Figure \ref{fig:FE2}a where each of the twelve bonds should be understood to have a different probability, and the shapes on the external vertices indicate how the base is embedded in the infinite lattice. Now when deleting and contracting a bond, we will not immediately see solvable lattices. Nevertheless, we can recursively apply the deletion-contraction algorithm until known lattices appear. The result of this is a binary tree of graphs, the size of which increases exponentially with the number of bonds in the base. For this reason, a computer is needed to handle large bases, but even then the maximum size achieved in this work is only 36 bonds. However, this is generally sufficient to provide estimates to thresholds within $10^{-7}$ of the numerical values. 

The $12^\mathrm{th}$-order polynomial for the $(4,8)$ base in Figure \ref{fig:FE2}a is \cite{Scullard11},
\begin{equation}
 1 - 4p^4 - 16p^6 + 12p^7 + 22p^8 + 16p^9 - 70p^{10} + 48p^{11} - 10p^{12} = 0, \label{eq:FE2}
\end{equation}
with solution $p_c= 0.67678736...$, differing from the numerical value by $1.5 \times 10^{-5}$ and cutting the error in half from the 6-bond base. At first glance, it appears that the generalized critical polynomial depends on the order in which bonds are chosen in the deletion-contraction algorithm. However, in \cite{Scullard11-2}, an argument was given that the generalized critical surface, and therefore the polynomial, is independent of the bond order. In the computer implementation, the answers are tested by performing the calculations several times using random bond orders.

\section{Implementation} \label{sec:impl}
The computer implementation of this algorithm is written in C++. As input, it takes a collection of bonds and a set of vertices, along with the number, $N_e$, of external vertices. Each bond and vertex is assigned a number, and the problem is fully specified once the two end vertices of each bond are given and identifications are made between external vertices to indicate how the base is embedded into the infinite lattice. Two vertices that are identified are called ``equivalent''. For a given lattice $L$, the solve routine works as follows:
\begin{enumerate}
 \item Choose a bond, $b$, on which to perform the deletion and contraction. A minimal set of requirements is used here. Of course, the same bond is chosen for both deletion and contraction.
 \item Delete $b$ to form lattice $L_0$, and contract the bond to form $L_1$.
 \item Simplify $L_0$ and $L_1$ by removing or combining any superfluous bonds that appeared as a result of deletion or contraction.
 \item Check if $L_0$ is one of the known lattices. If so, assign it the proper name and stop. If not, recursively call the solve routine for this lattice.
 \item Check if $L_1$ is one of the known lattices. If so, assign it the proper name and stop. If not, recursively call the solve routine for this lattice.
\end{enumerate}
This process produces a binary tree of lattices, with branches that terminate on identified graphs. The run-time is related to the number of lattices that appear in this tree. However, the population is determined by the speed with which known lattices appear, and only loosely related to the number of bonds, $n$, in the base. As different lattices and bases can have very different connectivity properties, the run-time may vary considerably for problems with the same $n$. But even for a given lattice, the choice of bond, $b$, at each step can significantly change the speed with which the solution is found, although the final result is independent of these choices. However, adding a bond to a base essentially doubles the number of lattices that will appear in the tree, and thus the number of operations increases as roughly $2^n$ for a given problem.
\subsection{Setup}
\begin{figure}
\begin{center}
\includegraphics{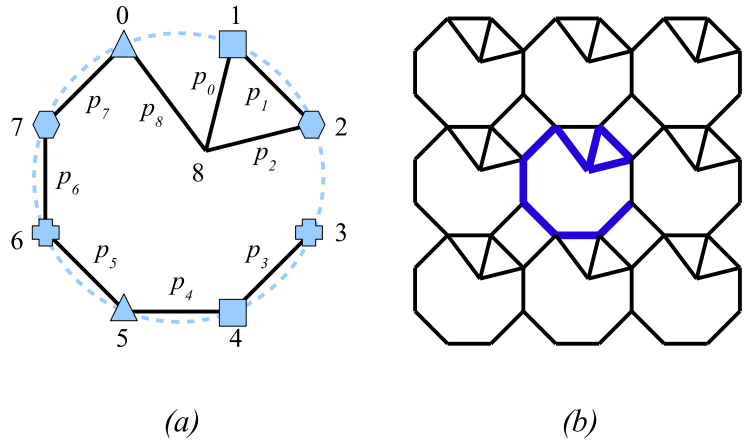}
\caption{a) A nine-bond base. The shapes indicate external vertex identifications. b) the lattice resulting from tiling this base.}
\label{fig:trioct}
\end{center}
\end{figure}

Figure \ref{fig:trioct} shows an example of a lattice partitioned into a base with nine bonds and nine vertices, eight of which are external. The embedding of the base is specified by identifying vertices that are equivalent in the infinite lattice and these identifications are illustrated by matching shapes in Figure \ref{fig:trioct}a. The numbering of the bonds and internal vertices is completely arbitrary. However, the external vertices are always assigned the first $N_e$ numbers and their ordering {\it is} important. Note that we only consider connected bases that can thus be contained inside a simple boundary loop that runs through all the external vertices in the obvious way (the dotted curve in Figure \ref{fig:trioct}a). The vertices are numbered counter-clockwise around this loop. When we delete bonds and simplify lattices, this ordering will be crucial in recognizing equivalent external gaps.
\subsection{Deletion} \label{sec:Deletion}
\begin{figure}
\begin{center}
\includegraphics{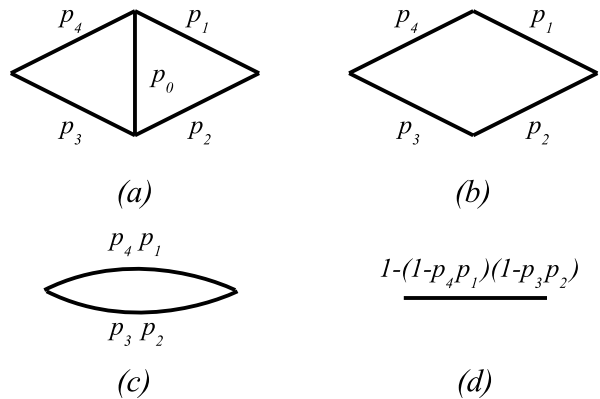}
\caption{Deletion of bond $0$ in a) puts bonds $1$ and $4$, and $2$ and $3$, in series, as depicted in b); c) simplifying these bonds leads to two bonds in parallel; d) the result of all the simplifications yields one bond with an effective probability.}
\label{fig:deletion}
\end{center}
\end{figure}

When a bond is deleted, it is simply removed from the list. Some simplification of the resulting lattice may be necessary. For example, we may create bonds doubled in series, like bonds $p_1$ and $p_4$, and $p_2$ and $p_3$, of Figure \ref{fig:deletion}b. These are replaced by single bonds with probabilities $p_1 p_4$ and $p_2 p_3$, as shown in Figure \ref{fig:deletion}c. Although series bonds are usually the only direct result of deletion, Figure \ref{fig:deletion}c indicates how simplifying these bonds can indirectly lead to bonds doubled in parallel. These bonds also need to be replaced with effective bonds with the appropriate probabilities. For example, in Figure \ref{fig:contraction}b, which is the result of contracting bond $0$ in Figure \ref{fig:contraction}a, bond $p_1$ is deleted, and $p_2$ is assigned the new probability
\begin{equation}
 1-(1-p_1)(1-p_2)=p_1+p_2-p_1 p_2
\end{equation}
which is simply the statement that in order for the effective single bond to be open, $p_1$ and $p_2$ cannot both be closed.
\subsection{Contraction}
\begin{figure}
\begin{center}
\includegraphics{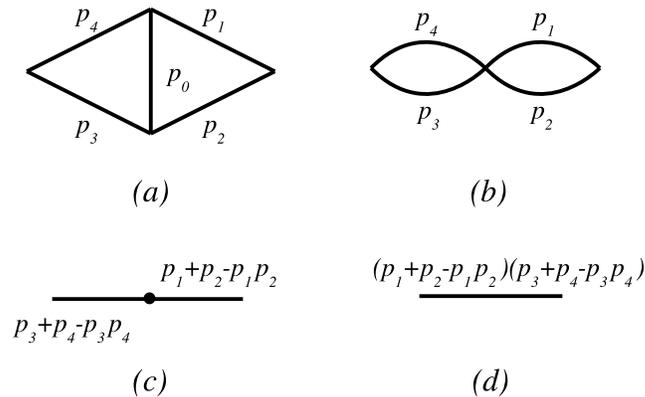}
\caption{Contraction of the bond $p_0$ in a) leads to doubled parallel bonds in b). Simplifying the parallel bonds leads to c) and then simplifying the series bonds gives d).}
\label{fig:contraction}
\end{center}
\end{figure}

When a bond is contracted, it is removed from the base, and its end vertices are merged into a single vertex. The direct result of this may be to create bonds doubled in parallel, as shown in Figure \ref{fig:contraction}b, but Figure \ref{fig:contraction}c illustrates how simplifying these parallel bonds can indirectly lead to bonds doubled in series. Further complications are possible here. Consider the base and lattice in Figure \ref{fig:trioct}. Contracting the $p_6$ bond in Figure \ref{fig:trioct}a necessitates the merger of vertices $6$ and $7$. However, examination of the embedding in Figure \ref{fig:trioct}b reveals that this gap is equivalent to the gap between vertices $2$ and $3$, and we must merge these vertices as well. The identification of equivalent gaps will be described in section \ref{sec:simplification}.
\subsection{Bond selection} \label{sec:bondselection}
\begin{figure}
\begin{center}
\includegraphics{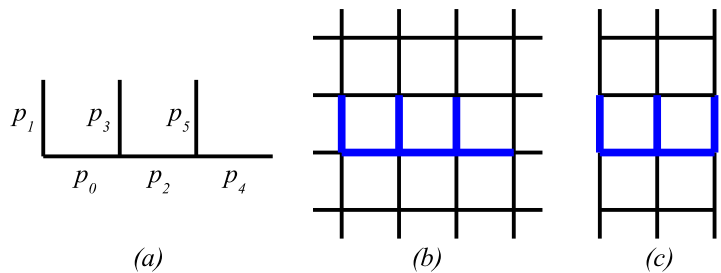}
\caption{a) a six-bond base for the square lattice; b) the embedding of this base; c) the lattice strip resulting from setting $p_4=0$.}
\label{fig:sqchannel}
\end{center}
\end{figure}

\begin{figure}
\begin{center}
\includegraphics{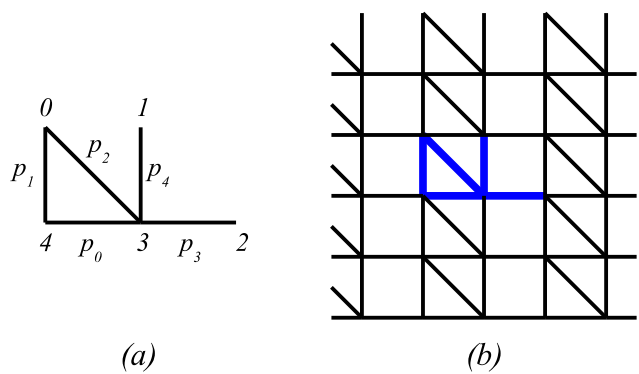}
\caption{a) a base for the $(3^3,4^2)$ lattice; b) its embedding. The bond $p_2$ is rejected by the algorithm because it joins external vertices that are not neighbours on the boundary loop of the base.}
\label{fig:crossbond}
\end{center}
\end{figure}

\begin{figure}
\begin{center}
\includegraphics{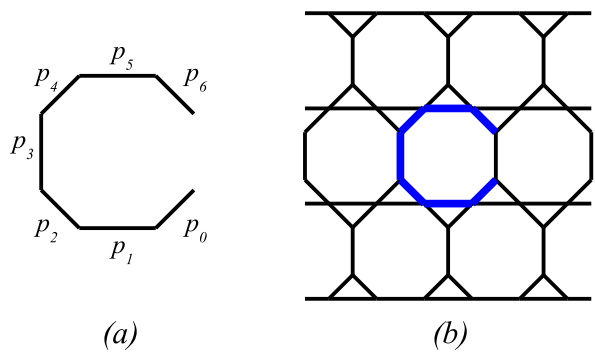}
\caption{a) a base for the kagome-like lattice in b). The bond selection algorithm laid out in section \ref{sec:bondselection} has no legal choices here.}
\label{fig:stopsign}
\end{center}
\end{figure}

Despite the earlier comments about the impact of bond selection on run time, I have not made any serious effort to optimize the bond choice for efficiency. Nevertheless, there is a variety of reasons a bond may be rejected in the present scheme:
\begin{enumerate}
\item Deleting it will disconnect the base. The algorithm is only used on bases with a single connected component. Note that, in some cases, contracting a bond may also result in a disconnected base. Consider the situation in Figure \ref{fig:trioct}. Contracting bond $p_0$ necessitates the removal of bond $p_4$ or $p_8$, since they will now span the same gap and need to be combined into a single bond. However, removing either of them disconnects the base. Before settling on a candidate bond, we must first examine the simplified lattices that result upon both deletion and contraction. It is possible to disconnect the lattice without disconnecting the base, with the result being a lattice strip, as shown in Figure \ref{fig:sqchannel}.
\item It is a ``supporting'' bond, that is, its endpoints are equivalent external vertices (e.g., any odd bond in Figure \ref{fig:sqchannel}a). Contracting such a bond collapses the lattice and can result in very complicated situations. It is certainly not impossible to deal with these, but I have avoided them for simplicity. The exception to this is the case of a lattice strip (Figure \ref{fig:sqchannel}c). As this is essentially a one-dimensional problem, it can only be critical when at least one of the supporting bonds has probability $1$. This means that if there is a supporting bond, $p_1$, the critical surface is of the form $(1-p_1) f=0$, where $f$ is some function of the other probabilities. Thus, the result of contracting this bond gives zero, bringing about a quick identification.
\item It connects two inequivalent external vertices that are not adjacent on the perimeter of the base, like bond $p_2$ in Figure \ref{fig:crossbond}. Contracting such a bond does not collapse the lattice, but nevertheless leads to complications that I chose to avoid.
\end{enumerate}
After settling on a set of rules, one may be concerned about the existence of lattices for which there is no legal bond choice, and in fact there are such lattices for these rules. An example is shown in Figure \ref{fig:stopsign}, which appropriately has the appearance of an array of stop signs. Other examples are of a similar nature. Although removing bonds $1$ to $5$ would clearly disconnect the base, it is not so obvious why $p_0$ or $p_6$ are bad choices. The reason is that contracting, say, $p_0$, makes $p_2$ and $p_5$ into parallel bonds through their external connections. But combining these into a single bond necessitates removing one of them, which disconnects the base. The bond $p_6$ has the same problem. There are several ways around this. One would be to re-partition the lattice into a different base that allows a good bond choice; there are clearly better bases to select for this kagome-like lattice. Another is to drop the restriction against disconnected bases. However, for bases of $36$ bonds, the maximum considered here, the appearance of such lattices is infrequent enough that I simply discard calculations in which they appear. In order to check that the final answer is correct, I run the algorithm several times for a given base anyway, as previously mentioned, each with a different labelling of the bonds to ensure a different path is taken.

\subsection{Simplification} \label{sec:simplification}
During the simplification phase, the algorithm checks the base for extraneous bonds or sites, iterating until no further changes need to be made. It seeks bonds doubled in series, bonds doubled in parallel, stranded vertices and dead ends. It also checks if a bond deletion has resulted in a lattice strip, which changes the bond selection rules slightly.
\subsubsection{Series bonds}
\begin{figure}
\begin{center}
\includegraphics{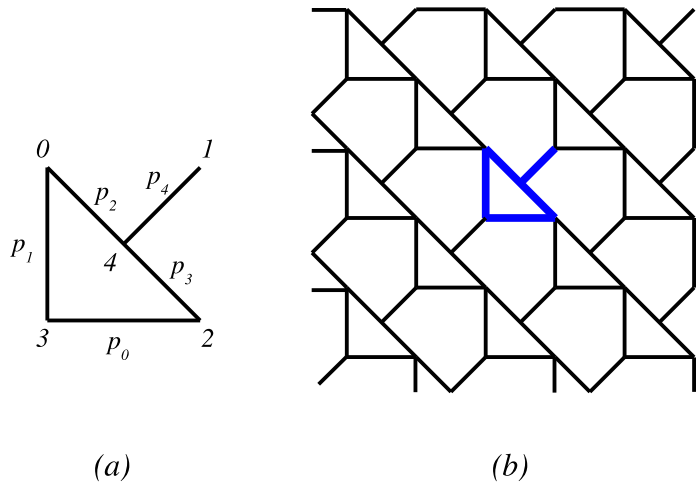}
\caption{The ``arrow'' lattice. Deletion of bond $0$ puts bonds $1$ and $4$ in series, but these are not simplified by the algorithm.}
\label{fig:arrow}
\end{center}
\end{figure}

If an internal vertex is incident with only two bonds, then those bonds are doubled in series, as in Figure \ref{fig:deletion}b, and they are replaced by a single bond, as shown in Figure \ref{fig:deletion}c. It is necessary that we consider only internal vertices here, as an external vertex between only two bonds is usually not indicative of bonds doubled in series. This can be seen in Figure \ref{fig:trioct}a, where, among other examples, vertex $6$ is between only bonds $p_5$ and $p_6$, but in the full lattice these bonds are not in series. In some cases, an external vertex connected with only one bond may indicate a bond doubled in series, which would happen for the lattice in Figure \ref{fig:arrow} if $p_0$ were deleted, but I do not replace these.
\subsubsection{Parallel bonds}
If two bonds have the same two end vertices, then they are doubled in parallel and one must be removed. This is done by deleting the bond with the lower number, and assigning the effective probability to the other.

\begin{figure}
\begin{center}
\includegraphics{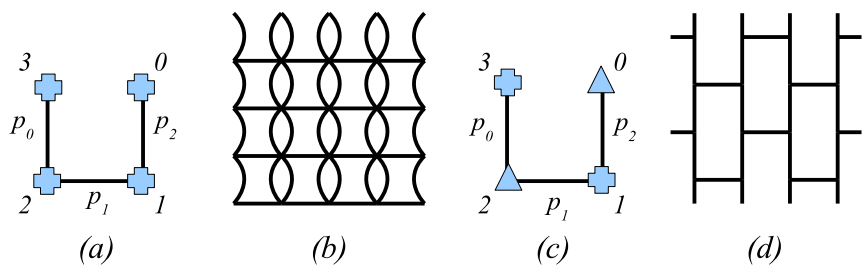}
\caption{The base in a) embeds to give the lattice b), and c) leads to d). In a), the bonds $p_0$ and $p_2$ span equivalent gaps, whereas in c) they do not.}
\label{fig:U}
\end{center}
\end{figure}

It is straightforward to find such parallel cases involving internal vertices, but when parallel bonds are between equivalent external vertices, special care is needed. Consider Figure \ref{fig:U}a, which shows a three-bond base with external vertex connections that lead to the lattice shown in Figure \ref{fig:U}b. Clearly, the bonds $p_0$ and $p_2$ span equivalent gaps, as the resulting lattice is just the square lattice with these two bonds doubled in parallel. However, Figure \ref{fig:U}c shows a base with the same configuration of bonds but different identifications of external vertices. Tiling this base gives the hexagonal lattice, shown in Figure \ref{fig:U}d in the ``brick wall'' representation, and now $p_0$ and $p_2$ do not span equivalent gaps, even though each end vertex of $p_0$ is equivalent with an endvertex of $p_2$. It is necessary to have some way to discriminate between these cases and it is here that the correct numbering of the external vertices becomes important. External vertices that neighbour on the boundary loop need not have a bond between them. However, we can think of their gap as a directed arc of the loop, $(v_1,v_2)$, oriented in the clockwise direction, with $v_1<v_2$. The gap between the last external vertex, $v_{N_e-1}$, and $v_0$ is written $(v_{N_e-1},v_0)$. In Figure \ref{fig:U}c, the bond $p_0$ spans the gap $(2,3)$. Vertex $2$ is equivalent to $0$ and vertex $3$ is equivalent to $1$, and substituting these in the gap $(2,3)$ gives $(0,1)$, which is the correctly oriented gap spanned by bond $p_0$. However, in the base of Figure \ref{fig:trioct}a, the bond $p_6$ bridges the gap $(6,7)$ with vertex $6$ equivalent with $3$ and $7$ with $2$. Substituting these into $(6,7)$ gives $(3,2)$, an arc directed counterclockwise, as $v_1>v_2$, which indicates that the gaps $(6,7)$ and $(3,7)$ are equivalent, as can be seen in Figure \ref{fig:trioct}b. More plainly, two external gaps are equivalent if the low vertex of one is equivalent with the high vertex of the other and vice versa. Note that in Figure \ref{fig:trioct}a, all vertices are equivalent so this condition is again met.

\subsubsection{Stranded vertices}
A stranded vertex is one which is not an endpoint for any bond. This can happen for external vertices, as discussed in section \ref{sec:Deletion}. Stranded vertices are easily found and removed. However, if another external vertex, $v$, was equivalent only with the stranded vertex, $w$, then $v$ is no longer external and must be demoted to internal as soon as $w$ is removed. For example, deleting $p_0$ and $p_1$ in Figure \ref{fig:arrow}b leaves vertex $3$ stranded. But vertex $1$ was equivalent only with vertex $3$ and thus the demotion of $1$ to internal must accompany the removal of $3$. This in turn leaves $1$ a dead-end vertex.
\subsubsection{Dead ends}
A dead end is an internal vertex that is incident with only one bond. This may occur upon demotion of an external vertex, as just described. A dead-end bond has no impact on the percolation process and is deleted along with the vertex.
\subsection{Identification}
The algorithm can identify a small number of simple graphs. Although the more lattices the algorithm is able to identify, the sooner the process terminates, it can be difficult to correctly identify graphs in which not all bonds are equivalent because a bond-matching procedure would need to be devised. As such, I only check for the square, triangular, hexagonal, and one-dimensional lattices. A lattice with three external vertices, three bonds, and one internal vertex is the hexagonal lattice; three external vertices, no internal, and three bonds is the triangular lattice; three external vertices and two bonds is the square lattice; and a single bond is one-dimensional. Furthermore, as previously described, if a strip lattice is collapsed, the result is zero. The capability to identify these graphs is sufficient to ensure that the algorithm always succeeds, provided it does not run into a lattice of the type in Figure \ref{fig:stopsign}.
\section{Results}
Now I list the results obtained by applying this method to various bases on the Archimedean lattices, except the kagome, which was already discussed in \cite{Scullard11-2}. For numerical results, we rely heavily on the paper of Parviainen \cite{Parviainen}, which, with the exception of $(3,12^2)$ for which Ding et al. \cite{Ding2010} provide a more accurate estimate, is still the standard reference for these lattices.
\subsection{$(4,8^2)$ lattice}
\begin{figure}
\begin{center}
\includegraphics{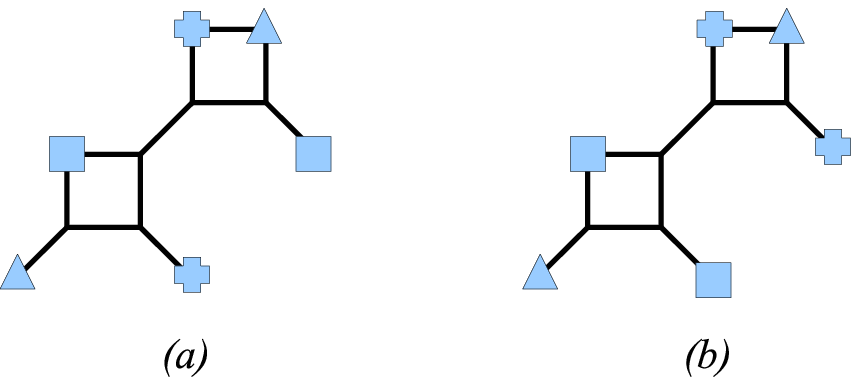}
\caption{Two different embeddings of the $(2 \times 1)$ base for the $(4,8^2)$ lattice.}
\label{fig:FE2}
\end{center}
\end{figure}

\begin{figure}
\begin{center}
\includegraphics{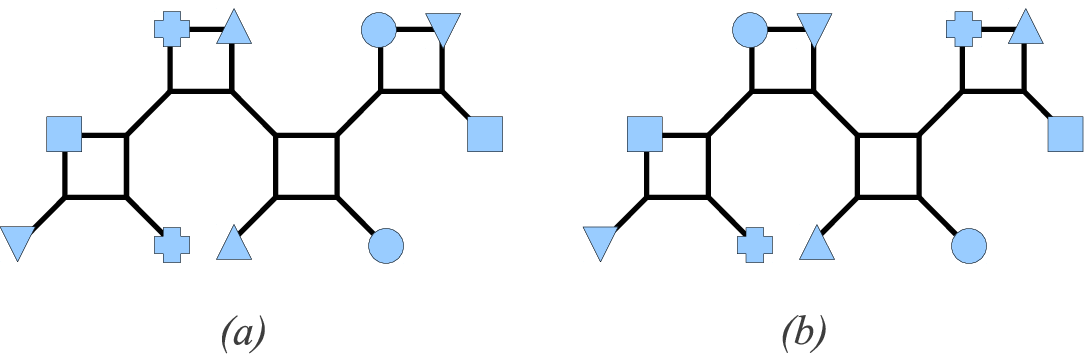}
\caption{Two different embeddings of the $(2 \times 2)$ base for the $(4,8^2)$ lattice.}
\label{fig:FE4}
\end{center}
\end{figure}

The single unit cell base result for this lattice was reported in \cite{Scullard08}, and the $12$-bond base consisting of two unit cells was studied in \cite{Scullard11}. In that work, however, only the base in Figure \ref{fig:FE2}a was used. Another option is to wire the external vertices as in Figure \ref{fig:FE2}b, but the result of this is the same as that of the $6$-bond base. Here, we will label different bases as $(N \times M)$ where $N$ and $M$ denote the number of unit cells in each direction. For example, the $(2 \times 2)$ base in Figure \ref{fig:FE4}a results in the polynomial
\begin{eqnarray}
 1 &-& 4 p^4 - 8 p^6 + 8 p^7 + 16 p^9 - 52 p^{10} 
+ 16 p^{11} - 52 p^{12} + 296 p^{13} \cr
&+& 160 p^{14} - 432 p^{15} - 1142 p^{16} + 712 p^{17} + 2436 p^{18} \cr
&-& 80 p^{19} - 7714 p^{20} + 10520 p^{21} - 6332 p^{22} + 1872 p^{23} - 222 p^{24},
\end{eqnarray}
predicting $p_c=0.67678965...$, which is slightly closer to the numerical value, $0.67680232(63)$, than the $12$-bond base, but is not a great improvement. Another base that might also be called $(2 \times 2)$ is shown in Figure \ref{fig:FE4}b, which is identical to Figure \ref{fig:FE4}a except for the identifications of external vertices. This change leads to a different polynomial,
\begin{eqnarray}
 1 &-& 16 p^6 + 4 p^8 - 84 p^{10} + 144 p^{11} + 88 p^{12} + 176 p^{13} \cr
&-& 300 p^{14} - 144 p^{15} - 1166 p^{16} + 1248 p^{17} + 968 p^{18} + 4440 p^{19} \cr
&-& 16394 p^{20} + 19392 p^{21} - 11264 p^{22} + 3296 p^{23} - 390 p^{24},
\end{eqnarray}
with $p_c=0.67681105...$, placing us within $8.73113 \times 10^{-6}$ of the numerical value.

\begin{figure}
\begin{center}
\includegraphics{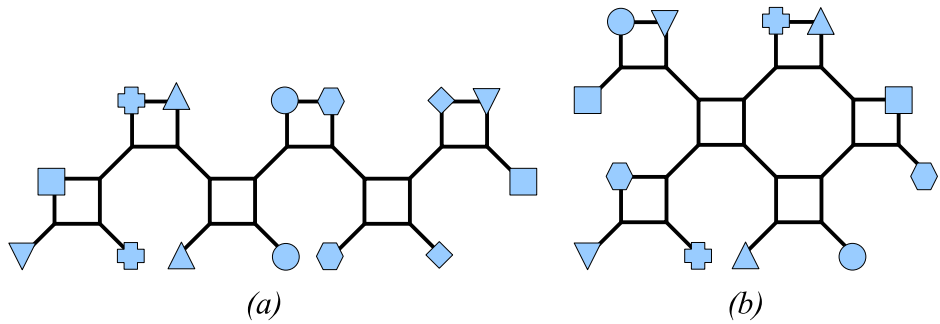}
\caption{$36-$bond bases for the $(4,8^2)$ lattice; a) $(2 \times 3)$, b) $(3 \times 2)$ . }
\label{fig:FE6}
\end{center}
\end{figure}

Moving to bases of $36$ bonds, two examples are given in Figures \ref{fig:FE6}a and \ref{fig:FE6}b, though many more exist. The first results in the polynomial,
\begin{eqnarray}
1 &-& 6 p^4 - 12 p^6 + 12 p^7 + 9 p^8 + 24 p^9 + 30 p^{10} - 24 p^{11} - 16 p^{12} \cr
&-& 108 p^{13} - 168 p^{14} + 12 p^{15} - 612 p^{16} + 1428 p^{17} + 1532 p^{18} \cr
&+& 3144 p^{19} - 5349 p^{20} - 15528 p^{21} - 3246 p^{22} + 41688 p^{23} + 40750 p^{24} \cr
&-& 77076 p^{25} - 127878 p^{26} + 143284 p^{27} + 243069 p^{28} - 234972 p^{29} \cr
&-& 607984 p^{30} + 1414836 p^{31} - 1365693 p^{32} + 758376 p^{33} - 253242 p^{34} \cr
&+& 47616 p^{35} - 3898 p^{36}
\end{eqnarray}
giving $p_c=0.6767896635...$, again closer to the numerical solution. However, the $(3 \times 2)$ base in Figure \ref{fig:FE6}b gives the polynomial
\begin{eqnarray}
 1 &-& 12 p^7 - 3 p^8 - 34 p^9 - 12 p^{10} + 6 p^{11} - 6 p^{12} - 36 p^{13} \cr
&+& 372 p^{14} + 386 p^{15} + 378 p^{16} - 156 p^{17} - 1248 p^{18} - 1800 p^{19} \cr
&-& 5655 p^{20} - 4388 p^{21} + 12528 p^{22} + 39930 p^{23} + 62136 p^{24} \cr
&-& 183384 p^{25} - 215964 p^{26} + 345672 p^{27} + 484599 p^{28} + 48816 p^{29} \cr
&-& 3588878 p^{30} + 6949548 p^{31} - 6640005 p^{32} + 3755348 p^{33} \cr
&-& 1284996 p^{34} + 247584 p^{35} - 20728 p^{36} 
\end{eqnarray}
with solution $p_c=0.67680215...$, a prediction that falls within the standard error of the numerical result, and is thus not ruled out. However, it should be stressed that no finite base will ever give the exact solution, as demonstrated in \cite{Scullard11-2}.
\begin{table}
\begin{center}
\begin{tabular}{clc}
base & & $p_c$ \\
\hline
$(2 \times 2)a$ &\vline& $0.67678965...$ \\
$(2 \times 2)b$ &\vline& $0.67681105...$ \\
$(2 \times 3)$ &\vline&  $0.67678966...$ \\
$(3 \times 2)$ &\vline&  $0.67680215...$ \\
\end{tabular}
\end{center}
\caption{Polynomial predictions for various bases of the $(4,8^2)$ lattice.}
\label{table:FE}
\end{table}
\subsection{$(3^3,4^2)$ lattice}
\begin{figure}
\begin{center}
\includegraphics{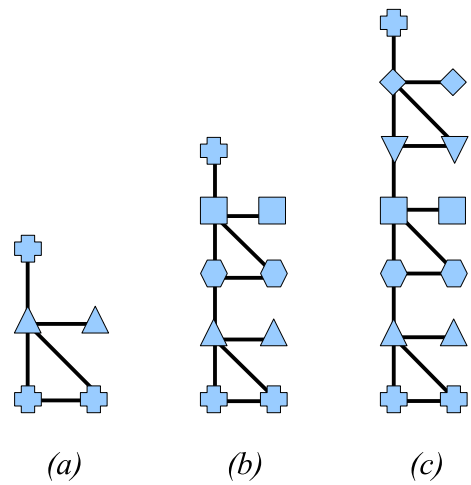}
\caption{$(N \times 1)$ bases for the $(3^3,4^2)$ lattice; a) $(1 \times 1)$; b) $(2 \times 1)$; c) $(3 \times 1)$. }
\label{fig:TCFSvert}
\end{center}
\end{figure}

\begin{figure}
\begin{center}
\includegraphics{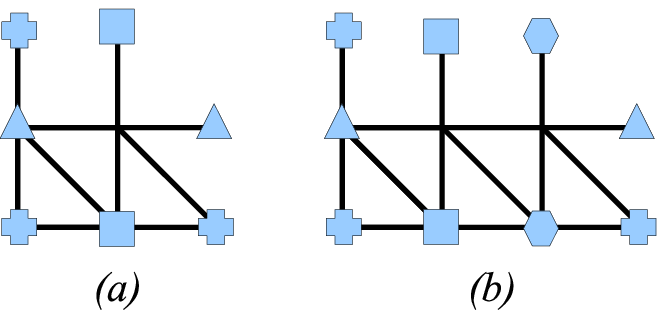}
\caption{$(1 \times N)$ bases for the $(3^3,4^2)$ lattice; a) $(1 \times 2)$; b) $(1 \times 3)$. }
\label{fig:TCFShorz}
\end{center}
\end{figure}

Although there is no finite base that will give the exact answer for any of these lattices, it is not clear in what manner the base must become infinite. In particular, it may be possible that a base need only be infinite in one direction, i.e. a strip. The results of this section will rule out this possibility for the $(3^3,4^2)$ lattice (Figure \ref{fig:archimedean}f). We will start with the progression shown in Figure \ref{fig:TCFSvert}, in which the bases consist of unit cells stacked in the vertical direction. One unit cell (Figure \ref{fig:TCFSvert}a) gives the polynomial reported in \cite{Scullard10},
\begin{equation}
 1 - 2 p - 2 p^2 + 3 p^3 - p^4 = 0 \label{eq:tcfs}
\end{equation}
with solution $p_c=0.419308168...$, fairly different from Parviainen's $p_c^{\mathrm{num}} = 0.41964191(43)$. Extending the base to the $(2 \times 1)$ configuration (Figure \ref{fig:TCFSvert}b) gives a polynomial that can be written in the factored form,
\begin{equation}
(1 - 2 p - 2 p^2 + 3 p^3 - p^4) (1 - 2 p + 2 p^2 + p^3 - p^4)=0
\end{equation}
and we recognize the first term in brackets as the polynomial (\ref{eq:tcfs}). The second term contains no root in $[0,1]$ and therefore the prediction is the same as for the $5-$bond base. Similarly, stacking three cells as in Figure \ref{fig:TCFSvert}c gives the factored form,
\begin{equation}
(1 - 2 p - 2 p^2 + 3 p^3 - p^4)(1 - p + p^2)(1 - 3 p + 2 p^2 + 4 p^3 - 2 p^4 - 2 p^5 + p^6)
\end{equation}
and once again we have the same prediction. It is probably safe to conjecture that this trend continues. The story is different if we extend the base in the horizontal direction. The polynomial for the $(1 \times 2)$ (Figure \ref{fig:TCFShorz}a) case was reported in \cite{Scullard10}, with the result $p_c=0.419614759...$ . For $(1 \times 3)$ (Figure \ref{fig:TCFShorz}b), we find,
\begin{eqnarray}
 1 &-& 3 p^2 - 8 p^3 - 15 p^4 - 3 p^5 + 220 p^6 + 84 p^7 - 2052 p^8 \cr
 &+& 4698 p^9 - 5343 p^{10} + 3471 p^{11} - 1231 p^{12} + 159 p^{13} \cr
&+& 30 p^{14} - 9 p^{15}=0
\end{eqnarray}
with solution $0.419650951...$ . The sequence for the remaining bases is shown in Table \ref{table:TCFS}. Clearly, these values are converging to a limit, which appears to be similar to the numerical value but not actually correct. In particular, the estimates for $n=5$, $6$, and $7$ indicate that the first seven digits are $0.4196551$, but Parviainen's result rules this out easily, as it seems extremely likely that the first five digits are in fact $0.41964$. This indicates that the $(1 \times \infty)$ base does not provide the exact answer. We can also rule out the possibility that we only need a wider strip, say ($2 \times \infty$). If this were exact, then so would be the case in which we deleted every horizontal and diagonal bond in the second row. Contracting the remaining vertical bonds in that row, we would necessarily find the same formula for ($1 \times \infty$) that we already showed was incorrect. Therefore, for this lattice, the only way to get the correct threshold is to consider the $(N \times M)$ base in which both $N$ and $M$ go to infinity. As such, we turn to the $(2 \times 2)$, $(3 \times 2)$, and $(2 \times 3)$ cases. For $(2 \times 2)$, the polynomial is,
\begin{eqnarray}
1 &-& 4 p^2 - 8 p^3 - 4 p^4 + 40 p^5 - 108 p^6 + 372 p^7 + 326 p^8 \cr
&-& 2640 p^9 - 4132 p^{10} + 40124 p^{11} - 101829 p^{12} + 145944 p^{13} \cr
&-& 134736 p^{14} + 82372 p^{15} - 32199 p^{16} + 6904 p^{17} - 236 p^{18} \cr
&-& 224 p^{19} + 36 p^{20}=0
\end{eqnarray}
with solution $p_c=0.4196154184...$, barely distinguishable from, but still slightly better than, the $(1 \times 2)$ answer. The $(3 \times 2)$ base yields the polynomial
\begin{eqnarray}
 1 &-& 6 p^2 - 12 p^3 + 6 p^4 + 132 p^5 - 58 p^6 - 204 p^7 - 1275 p^8 \cr
&+& 1272 p^9 + 8514 p^{10} + 1836 p^{11} - 82380 p^{12} + 55062 p^{13} \cr
&+& 504828 p^{14} - 903004 p^{15} - 3039471 p^{16} + 18382050 p^{17} \cr
&-& 48094255 p^{18} + 83086308 p^{19} - 104909466 p^{20} + 100560770 p^{21} \cr
&-& 74085918 p^{22} + 41775876 p^{23} - 17645241 p^{24} + 5303532 p^{25} \cr
&-& 991518 p^{26} + 54936 p^{27} + 23244 p^{28} - 6048 p^{29} + 488 p^{30}
\end{eqnarray}
predicting $p_c=0.4196154196...$, which is hardly different from $(2 \times 2)$. The best estimate is given by the $(2 \times 3)$ base
\begin{eqnarray}
 1 &-& 4 p^3 - 21 p^4 - 72 p^5 - 32 p^6 + 42 p^7 + 1017 p^8 + 6160 p^9 \cr
&-& 5163 p^{10} - 49410 p^{11} - 78554 p^{12} + 466284 p^{13} + 1142880 p^{14} \cr
&-& 6161272 p^{15} - 1697280 p^{16} + 66144060 p^{17} - 221367606 p^{18} \cr
&+& 428244768 p{19} - 574288983 p^{20} + 567883664 p^{21} - 423773169 p^{22} \cr
&+& 239027052 p^{23} - 100024955 p^{24} + 29504220 p^{25} - 5317833 p^{26} \cr
&+&  236622 p^{27} + 138600 p^{28} - 33642 p^{29} + 2625 p^{30}
\end{eqnarray}
giving $p_c=0.4196453185...$, differing from the numerical result by $3.6 \times 10^{-6}$.
\begin{figure}
\begin{center}
\includegraphics{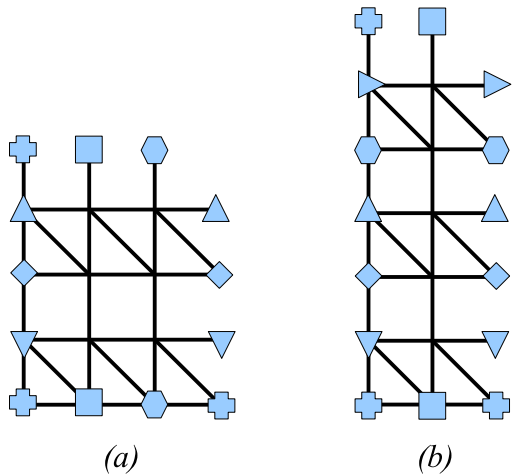}
\caption{a) $(2 \times 3)$ and b) $(3 \times 2)$ bases for the $(3^3,4^2)$ lattice. }
\label{fig:TCFSMxN}
\end{center}
\end{figure}
\begin{table}
\begin{center}
\begin{tabular}{clc}
$N$ & & $p_c$ \\
\hline
$1$ &\vline& $0.419308168...$ \\
$2$ &\vline& $0.419614759...$ \\
$3$ &\vline& $0.419650951...$ \\
$4$ &\vline& $0.419654761...$ \\
$5$ &\vline& $0.419655145...$ \\
$6$ &\vline& $0.419655183...$ \\
$7$ &\vline& $0.419655187...$ \\
\end{tabular}
\end{center}
\caption{Roots of generalized critical polynomials for the $(1 \times N)$ bases of the $(3^3,4^2)$ lattice. These appear to converge to a number ruled out by simulations.}
\label{table:TCFS}
\end{table}
\begin{table}
\begin{center}
\begin{tabular}{clc}
base & & $p_c$ \\
\hline
$(2 \times 2)$ &\vline& $0.4196154184...$ \\
$(3 \times 2)$ &\vline& $0.4196154196...$ \\
$(2 \times 3)$ &\vline& $0.4196453185...$ \\
\end{tabular}
\end{center}
\caption{Threshold predictions for various $(3^3,4^2)$ bases.}
\label{table:TCFS2}
\end{table}
\subsection{$(3,12^2)$ lattice}
\begin{figure}
\begin{center}
\includegraphics{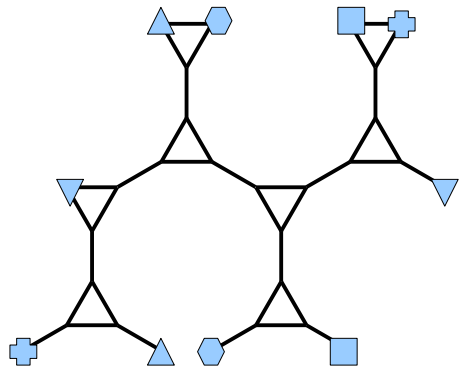}
\caption{$(2 \times 2)$ base for the $(3,12^2)$ lattice.}
\label{fig:3-12}
\end{center}
\end{figure}

The single-cell polynomial for the $(3,12^2)$ lattice has been reported in various places \cite{Scullard10}, but I note it here,
\begin{equation}
1 - 3 p^4 - 6 p^5 + 3 p^6 + 15 p^7 - 15 p^8 + 4 p^9=0 .
\end{equation}
Oddly enough, this polynomial can be written in factored form,
\begin{equation}
 (1 + p - 2 p^3 + p^4) (1 - p + p^2 + p^3 - 7 p^4 + 4 p^5)=0 .
\end{equation}
The second term in brackets is the one that has the root in $[0,1]$, $p_c=0.74042331...$, and therefore, for some reason, the prediction for this nine-bond base is given by a fifth order polynomial. The numerical value given by Parviainen is $p_c^{\mathrm{num}}=0.74042195(80)$, whereas Ding et al. \cite{Ding2010} give the more recent result $0.74042077(2)$. Turning to the $(2 \times 2)$ case (Figure \ref{fig:3-12}), we have the polynomial,
\begin{eqnarray}
 1 &-& 6 p^8 - 24 p^9 - 12 p^{10} + 48 p^{11} + 18 p^{12} - 120 p^{13}- 120 p^{14} \cr
&+& 252 p^{15} + 639 p^{16} - 192 p^{17} - 720 p^{18} + 804 p^{19} - 1302 p^{20} \cr
&-& 1024 p^{21} + 468 p^{22} - 14652 p^{23} + 28635 p^{24} + 79392 p^{25} \cr
&-& 181344 p^{26} - 88368 p^{27} + 425721 p^{28} - 17280 p^{29} - 996084 p^{30} \cr
&+& 1558008 p^{31} - 1244067 p^{32} + 603008 p^{33} - 180096 p^{34} \cr
&+& 30720 p^{35} - 2304 p^{36} =0
\end{eqnarray}
which does not factor, and predicts $p_c=0.74042099...$, a difference of $9.6 \times 10^{-7}$ from Parviainen's value, but within $2.2 \times 10^{-7}$ of Ding et al. This also compares favourably with the approximation given by Ziff and Gu, $p_c \approx 0.74042081$, based on a numerically fit universal condition. For the moment, this is as far as we can go with this lattice.
\subsection{$(4,6,12)$ lattice}
\begin{figure}
\begin{center}
\includegraphics{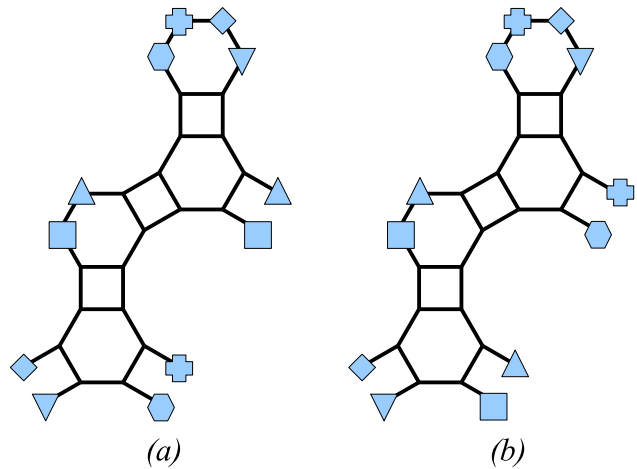}
\caption{Two $36$-bond bases for the $(4,6,12)$ lattice. These have the same critical polynomial.}
\label{fig:FST}
\end{center}
\end{figure}

The single unit cell polynomials for the $(4,6,12)$ and $(3^2,4,3,4)$ lattices were given in \cite{Scullard11}. Because of the size of their bases, we can only extend these estimates to two unit cells. Two possible bases for the $(4,6,12)$ lattice are shown in Figures \ref{fig:FST}a and \ref{fig:FST}b. These turn out to have the same polynomial,
\begin{eqnarray}
1 &-& 12 p^6 - 4 p^8 + 16 p^9 - 14 p^{10} + 32 p^{11} - 106 p^{12} \cr
&+& 72 p^{13} - 248 p^{14} + 680 p^{15} - 220 p^{16} + 2080 p^{17} - 2414 p^{18} \cr
&-& 1944 p^{19} - 5573 p^{20} + 6904 p^{21} - 1924 p^{22} + 24028 p^{23} \cr
&+& 21869 p^{24} - 48140 p^{25} - 201044 p^{26} + 261640 p^{27} + 198914 p^{28} \cr
&+& 148952 p^{29} - 2204780 p^{30} + 4153808 p^{31} - 3985112 p^{32} \cr
&+& 2288132 p^{33} - 800476 p^{34} + 158616 p^{35} - 13734 p^{36}
\end{eqnarray}
with solution, $p_c=0.69375829...$, an incremental improvement over the single-cell value, $p_c=0.69377849...$ reported in \cite{Scullard11}, but within $2.4 \times 10^{-5}$ Parviainen's numerical result, $p_c^{\mathrm{num}}= 0.69373383(72)$ .
\subsection{$(3^2,4,3,4)$ lattice}
\begin{figure}
\begin{center}
\includegraphics{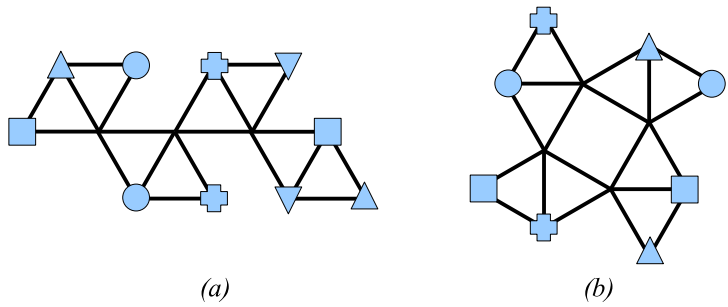}
\caption{$20$-bond bases for the $(3^2,4,3,4)$ lattice.}
\label{fig:TSFTF}
\end{center}
\end{figure}

Using the base in Figure \ref{fig:TSFTF}a, we find
\begin{eqnarray}
1 &-& 4 p^2 - 8 p^3 + 20 p^5 - 84 p^6 + 248 p^7 + 970 p^8 \cr
&-& 3748 p^9 -  4042 p^{10} + 41664 p^{11} - 101482 p^{12} \cr
&+& 139104 p^{13} - 122372 p^{14} + 70604 p^{15} - 25368 p^{16} \cr
&+& 4484 p^{17} + 234 p^{18} - 256 p^{19} + 34 p^{20}=0
\end{eqnarray}
giving $p_c=0.41412438...$ , a difference of $1.3 \times 10^{-5}$ from Parviainen's $0.41413743$, but his standard error is $4.6 \times 10^{-7}$. The single unit-cell prediction found in \cite{Scullard10} was $0.414120304...$, so we have improved on this estimate somewhat. We can do better with the base shown in Figure \ref{fig:TSFTF}b, for which we find the polynomial
\begin{eqnarray}
 1 &-& 8 p^3 - 36 p^4 - 44 p^5 + 88 p^6 + 912 p^7 + 266 p^8 - 7604 p^9 \cr 
 &-& 1390 p^{10} + 71480 p^{11} - 194138 p^{12} + 277116 p^{13} - 247748 p^{14} \cr 
 &+& 142984 p^{15} - 50264 p^{16} + 7940 p^{17} + 1010 p^{18} - 648 p^{19} + 82 p^{20},
\end{eqnarray}
and $p_c=0.41414477...$, putting us within $7.3 \times 10^{-6}$ of the numerical result.
\subsection{$(3^4,6)$ lattice}
\begin{figure}
\begin{center}
\includegraphics{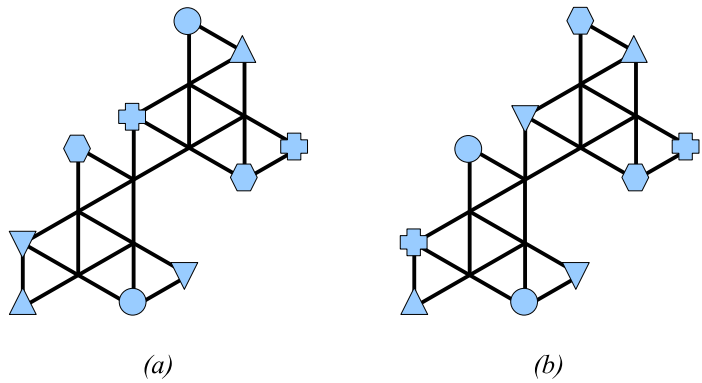}
\caption{$30$-bond bases for the $(3^4,6)$ lattice. These yield the same polynomial.}
\label{fig:TFS}
\end{center}
\end{figure}

The single-cell polynomial for the $(3^4,6)$ lattice was reported in \cite{Scullard11} with the approximation $p_c=0.43437077...$ . Extending to $30$ bonds, as shown in Figures \ref{fig:TFS}a and b, we have, for both these embeddings,
\begin{eqnarray}
 1 &-& 8 p^3 - 24 p^4 + 2 p^5 + 58 p^6 - 186 p^7 - 244 p^8 + 4100 p^9 \cr
&+& 7053 p^{10} - 41066 p^{11} - 90261 p^{12} + 353308 p^{13} + 817651 p^{14} \cr
&-& 4135392 p^{15} - 883466 p^{16} + 41247262 p^{17} - 138495809 p^{18} \cr
&+& 271182796 p^{19} - 372758548 p^{20} + 383898312 p^{21} - 304802391 p^{22} \cr
&+& 188649862 p^{23} - 90989211 p^{24} + 33847422 p^{25} - 9501925 p^{26} \cr
&+& 1938036 p^{27} - 268902 p^{28} + 22392 p^{29} - 823 p^{30}
\end{eqnarray}
with $p_c=0.43435240...$ compared with the numerical result $p_c^{\mathrm{num}}=0.43430621(50)$. The difference is $4.6 \times 10^{-5}$.
\subsection{$(3,4,6,4)$ lattice}
\begin{figure}
\begin{center}
\includegraphics{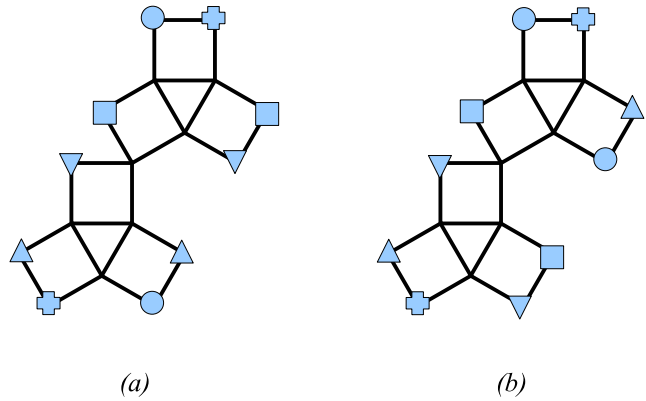}
\caption{Bases for the $(3,4,6,4)$ lattice. These have the same generalized critical polynomial.}
\label{fig:3464}
\end{center}
\end{figure}

The single-cell 12-bond polynomial for the $(3,4,6,4)$ lattice was reported in \cite{Scullard08,Scullard10}, which made the prediction $p_c=0.524821...$. The two $24$-bond bases in Figures \ref{fig:3464}a and b yield the polynomial
\begin{eqnarray}
 1 &-& 4 p^3 - 8 p^4 - 8 p^5 + 18 p^6 - 20 p^7 - 126 p^8 + 516 p^9 \cr
&+& 754 p^{10} - 1180 p^{11} - 6071 p^{12} + 976 p^{13} + 34222 p^{14} \cr
&+& 768 p^{15} - 257049 p^{16} + 652088 p^{17} - 866150 p^{18} + 730808 p^{19} \cr
&-& 412380 p^{20} + 155436 p^{21} - 37456 p^{22} + 5168 p^{23} - 304 p^{24}
\end{eqnarray}
with root $p_c=0.52483166...$, differing from Parviainen's $0.52483258(53)$ by only $9.1 \times 10^{-7}$.
\begin{table}
\begin{center}
\begin{tabular}{clclclc}
lattice & & $p_c^{\mathrm{poly}}$ & & $p_c^{\mathrm{num}}$ & & $|p_c^{\mathrm{poly}}-p_c^{\mathrm{num}}|$ \\
\hline
$(4,8^2)$ &\vline& $0.67680215...$ &\vline& $0.67680232(63)$ &\vline& --- \\
$(3^3,4^2)$ &\vline& $0.41964531...$ &\vline& $0.41964191(43)$ &\vline& $3.6 \times 10^{-6}$\\
$(3,12^2)$ &\vline& $0.74042099...$ &\vline& $0.74042077(2)$ &\vline& $2.2 \times 10^{-7}$\\
$(4,6,12)$ &\vline& $0.69375829...$ &\vline& $0.69373383(72)$ &\vline& $2.4 \times 10^{-5}$\\
$(3^2,4,3,4)$ &\vline& $0.41414477...$ &\vline& $0.41413743(46)$ &\vline& $7.3 \times 10^{-6}$ \\
$(3^4,6)$ &\vline& $0.43435240...$ &\vline& $0.43430621(50)$ &\vline& $4.6 \times 10^{-5}$ \\
$(3,4,6,4)$ &\vline& $0.52483166...$ &\vline& $0.52483258(53)$ &\vline& $9.1 \times 10^{-7}$\\
\end{tabular}
\end{center}
\caption{The best polynomial estimates for the Archimedean lattices along with the numerical values. All numerics are from \cite{Parviainen} except $(3,12^2)$, which is from \cite{Ding2010}.}
\label{table:bestpc}
\end{table}
\section{Conclusions}
I have presented the computation of generalized critical polynomials using a computer program to perform the deletion-contraction algorithm. The results clearly support the conjecture that these polynomials predict thresholds converging to the exact answer in the limit of an infinite base. Moreover, this seems to suggest that critical bond thresholds for most unsolved problems are not algebraic numbers, since no finite base can give the correct threshold unless the single-cell answer is exact. A summary of the best results for each lattice is in Table \ref{table:bestpc}. Their accuracy ranges from $4.6 \times 10^{-5}$ for the $(3^4,6)$ lattice, to the result for the $(4,8^2)$ lattice which is not ruled out by the numerical answer. Many paths for future work are available. One might hope that far greater accuracy is possible if a method can be found to compute the polynomials more efficiently than by using deletion-contraction and indeed this is a subject of current study. It should be noted that the generalized critical polynomials presented here are related to F.Y. Wu's homogeneity approximation \cite{Wu10} for the $q-$state Potts model \cite{Wu82}, which he has so far been able to apply to get a 6-bond kagome \cite{Wu79} polynomial and, in recent work with W. Guo \cite{Wu2012}, a 6-bond $(4,8^2)$ polynomial. In particular, his method predicts (\ref{eq:FE1}) in the $q \rightarrow 1$ limit. Recently, the deletion-contraction method was also generalized to the $q-$state Potts model \cite{Jacobsen12}, which allows accurate determination of critical points for large bases and arbitrary $q$, including predictions in the unphysical anti-ferromagnetic region. Finally, polynomials can be defined in three dimensions with deletion-contraction in exactly the same way as in two, but it remains to be seen if they can give any information about the critical point.

This work was performed under the auspices of the U.S. Department of Energy by
Lawrence Livermore National Laboratory under Contract DE-AC52-07NA27344.



\section*{References}

\bibliography{scullard}







\end{document}